# Mount Rainier Elevation Survey 2024

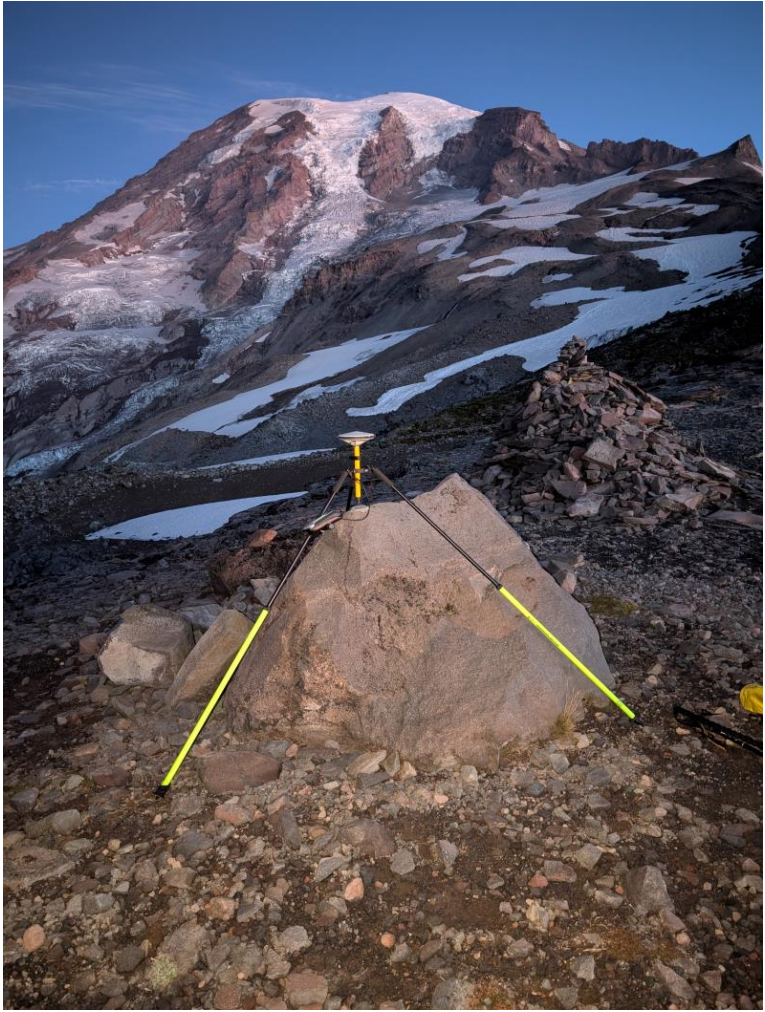

Mount Rainier as viewed from the McClure Rock monument, Mount Rainier National Park, Washington.
    Mt Rainier as viewed from McClure Rock monument

© Eric Gilbertson

Eric W Gilbertson[1], Larry Signani[2], Kathryn E Stanchak[3]

[1] Seattle University, Seattle, WA

[2] Enumclaw, WA

[3] Issaquah, WA

# Abstract

The elevation of Mount Rainier was last surveyed in 2010 by the Land Surveyors Association of Washington. More recent LiDAR data and observational reports have indicated that the historical highest point on the mountain, Columbia Crest, has lost a significant amount of elevation and may no longer be the highest point.

This report documents the results of multiple ground surveys conducted in August and September 2024 to determine the current elevation of Columbia Crest and the current elevation and location of the summit of Mt Rainier.

On August 28, 2024, a Promark 220 differential GPS unit was used to measure the elevation of Columbia Crest and the Southwest Rim, two recognized historical local maxima on the rim of the summit caldera. Columbia Crest is a year-round ice-capped locality, while the Southwest Rim melts to rock each summer. Measurements were corroborated with Abney level measurements and LiDAR data from 2007 and 2022.

On September 21 and 22, 2024, additional ground measurements were taken with the Promark 220 of the Summit 2 USGS monument location and of the highest rock on the Southwest Rim. Additionally, the three USGS monuments lower on the mountain (Muir, McClure, and Paradise) were measured. These measurements were used to corroborate the summit measurements.

Monument measurements were found to be consistent with 2010 measurements. Southwest Rim summit measurements were consistent from August and September. Columbia Crest, last measured at 14,411.0ft (NGVD29) in 1998, was measured to have lost 21.8ft of elevation. Now the ice-capped Columbia Crest is no longer the summit of Mt Rainier. The summit is now on the seasonally rocky Southwest Rim, next to the Summit 2 USGS monument location. (The physical monument was not located and has likely been stolen).

The new summit elevation of Mount Rainier is 14,399.6ft NGVD29 (14,406.2ft NAVD88).

# List of Terms or Acronyms

**NGVD29:** National Geodetic Vertical Datum of 1929

**NAVD88:** North American Vertical Datum of 1988

**LiDAR:** Light Detection and Ranging

**LSAW:** Land Surveyors Association of Washington

**USGS:** United States Geological Survey

**GPS:** Global Positioning System

**WSRN:** Washington State Reference Network

**RMI:** Rainier Mountaineering Institute

**OPUS:** Online Positioning User Service

**CSRS-PPP:** Canadian Spatial Reference System Precise Point Positioning processing

# Acknowledgements

Funding was provided by the American Alpine Club, with equipment provided by Seattle University. Surveyors from the 1988, 1998, and 2010 LSAW Rainier Survey teams helped plan methodology and processed results. Ross Wallette, Josh Spitzberg, Alden Ryno, and Saulius Braciulis helped conduct ground surveys. Dustin W provided summit pictures for photographic analysis. Kyle B helped with LiDAR data processing. Rainier climbing rangers and guides kept the Disappointment Cleaver route open so we could make it up to the summit for the first measurement in August. IMG guide Justin Sackett provided valuable information so we could reach the summit for the second measurement in September after ladders had been pulled on the mountain.

.

# Introduction

Mount Rainier is the tallest peak in Washington State, the most topographically prominent peak in the contiguous United States (US), and the most heavily glaciated peak in the contiguous US. Until recently, it was one of the few peaks in the contiguous US with a permanent icecap on the summit.

The elevation of such a significant peak is of interest to many people, including geologists, volcanologists, staff and visitors of Mount Rainier National Park, and the many mountaineers who attempt to reach the summit each year. Surveyors have been measuring the height of Mount Rainier since the mid-1800s (see Table 1). Early measurements using barometers brought to the summit were prone to large errors, but later triangulation-based measurements were more accurate. The triangulation method involves pointing a theodolite at the summit from a location of known elevation and position. An angle is measured to the summit and, using the angle, distance, and trigonometry, the summit elevation can be calculated. Measurements can be taken from multiple locations and the results averaged to give a final summit elevation.

It is important that the measurement of a peak like Mount Rainier be taken at the appropriate time of year. For a peak with a permanent icecap on the summit, the accepted elevation is the elevation of the icecap at the lowest snow time of year. This is generally late summer, when the seasonal snow from the last winter has melted as much as it will but autumn snows have not yet started accumulating. Measuring at this time of year ensures seasonal snow does not count towards the summit elevation.

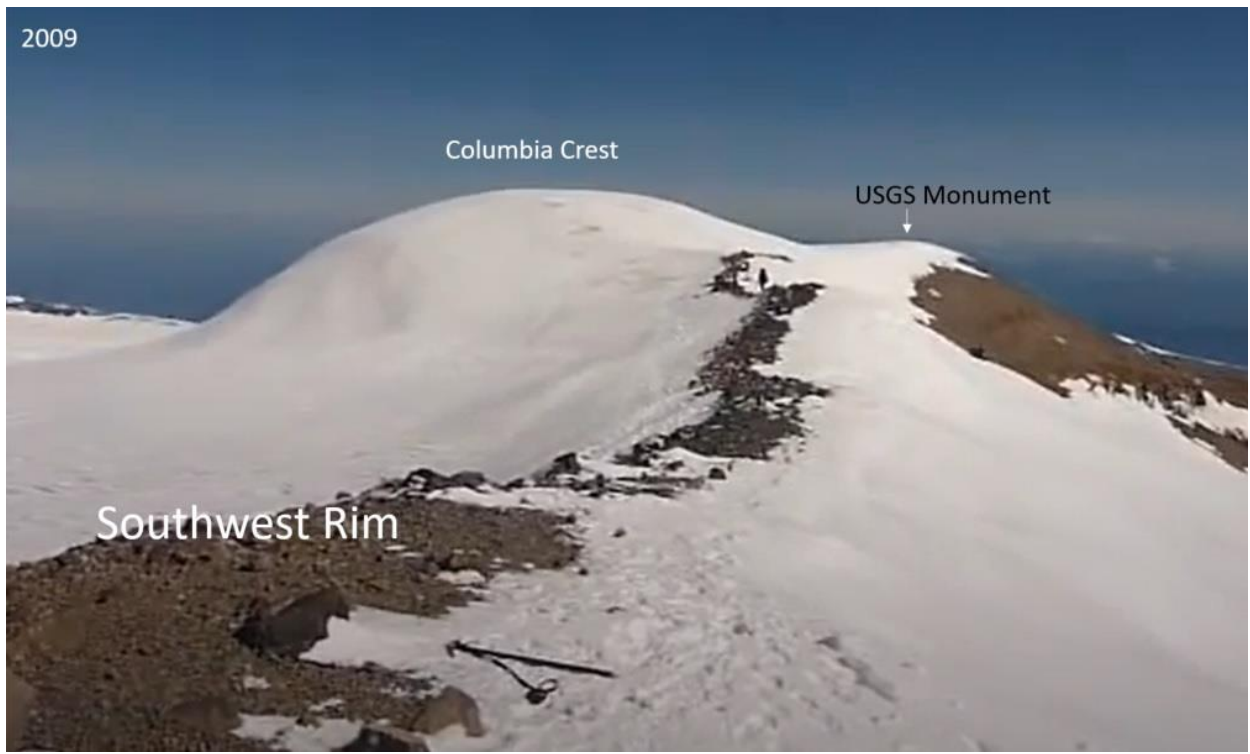

**Figure 1.** The view of Columbia Crest from the Southwest Rim in 2009 (Arhuber 2009).

Figure 1. Photograph taken in 2009 from the SW Rim looking towards Columbia Crest showing Columbia Crest.

The summit area of Mount Rainier has a crater rim that partially melts down to rock every summer, but there has historically been a permanent dome of ice on the west edge of the rim. This ice dome is referred to as Columbia Crest, and has historically been the highest point of the mountain (Fig. 1). Thus, the elevation of the highest point of ice on Columbia Crest in late summer has historically been considered the elevation of Mount Rainier.

The triangulation method was employed in the summer of 1914 and again in 1956 by the USGS (United States Geological Survey) to measure the elevation of Columbia Crest (Signani 2000). The 1956 survey measured the elevation 14,410 ft, and this is the elevation printed on the official USGS topographic maps (the quads; Fig. 2). This is also the elevation used by Mt Rainier National Park (National Park Service 2024). Note that this elevation is in the NGVD29 vertical datum. All elevations in this report will be reported in the NGVD29 datum unless otherwise noted so they can be fairly compared. Simply put, a datum is a reference point that surveyors use to define mean sea level extended across land. Thus, datums defined different ways can result in different elevations for the same point.

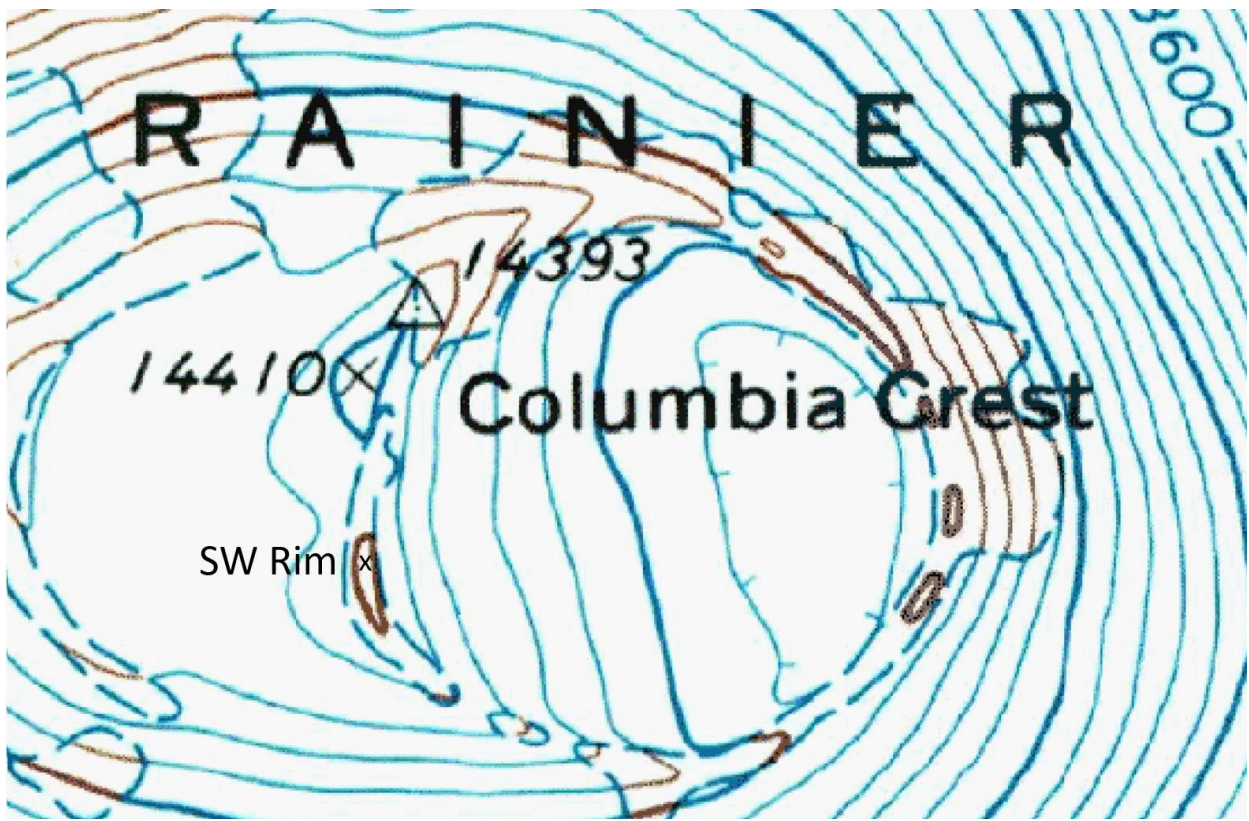

**Figure 2.** The summit region of Mount Rainier from the 1971 USGS quad (USGS 1971), with Columbia Crest and the Southwest Rim labeled.

Figure 2. Topographic map of the summit region of Mt Rainier showing Columbia Crest and the SW Rim.

Since 1988, several surveys of the elevation of Mount Rainier have been performed using GPS (Global Positioning System) technology to increase accuracy. This involves bringing a GPS unit to the summit and collecting data from many satellites to measure the summit elevation. In July, 1988, surveyors from the Land Surveyors Association of Washington (LSAW) mounted a GPS on top of Columbia Crest and measured an elevation of 14,411.1 ft (Signani 2000). They also measured the Summit 2 USGS monument on the Southwest Rim.

On August 27, 1998, surveyors from LSAW conducted another GPS survey finding almost the same summit elevation for Columbia Crest, 14,411.0 ft (Signani 2000). On July 23, 2010, LSAW surveyors returned and measured the USGS monuments around the mountain and in the summit area, though they did not measure Columbia Crest (Schrock 2011).

**Table 1**: Mount Rainier surveys over time (Matamoros 2006). Results given in ft, NGVD29.

| Year | Elevation (Columbia Crest, ft) | Elevation (SW Rim, ft) | Credit | Method | Date | Error (+/-ft) 95% confidence |
|------|-------------------------------|------------------------|--------|--------|------|------------------------------|
| 1841 | 12,333.0 | - | Lt Charles Wilkers | Triangulation | - | - |
| 1879 | 14,444.0 | - | James Smyth Lawson | Barometer | - | - |
| 1888 | 14,524.0 | - | E.S. Ingraham | Barometer | - | - |
| 1896 | 14,519.0 | - | USGS | Barometer | - | - |
| 1897 | 14,528.0 | - | Edgar McClure | Barometer | - | - |
| 1902 | 14,363.0 | - | USGS | Barometer | - | - |
| 1905 | 14,394.0 | - | Alexander McAdie | Barometer | - | - |
| 1914 | 14,408.0 | - | USGS | Triangulation | - | - |
| 1956 | 14,410.0 | - | USGS | Triangulation | - | - |
| 1988 | 14,411.1 | 14,401.5 | LSAW | GPS | July | 0.1 |
| 1998 | 14,411.0 | - | LSAW | GPS | Aug 27 | 0.1 |
| 2007 | 14,405.6 | 14,399.5 | USGS | LiDAR | Sept | 0.3 |
| 2010 | - | 14,400.7 | LSAW | GPS | July 23 | 0.1 |
| 2022 | 14,392.3 | 14,398.7 | USGS | LiDAR | | 0.3 |
| 2023 | 14,390.7 | 14,398.7 | Gilbertson | Photographic | July 23 | 0.9 |
| 2024 | 14,389.2 | 14399.6 | Gilbertson | dGPS | Aug 28 | 0.1 |
| 2024 | - | 14399.6 | Gilbertson | dGPS | Sept 21 | 0.1 |

The GPS units used in these more recent surveys are much more accurate than handheld consumer-grade GPS units, like those found in a phone. Consumer-grade units can have very high vertical errors, up to +/-20 ft or more, as a result of effects like atmospheric distortion, multipath errors, and receiving signals from only a limited number of available satellites.

Survey-grade differential GPS (dGPS) units can get errors down to +/-0.1 ft or better. They generally have access to more satellites, have external antennas to help reduce multipath errors, and are capable of correcting for atmospheric distortions using nearby base stations. Data usually needs to be collected over a long period of time, then post-processed. The US government provides a publicly-available software tool, OPUS (Online Positioning User Service) to process dGPS data (OPUS 2024). In Washington, the Washington State Reference Network (WSRN) can also be used to process dGPS data (WSRN 2024).

There was one other summit measurement taken in 2007 (USGS 2007) by a different method, LiDAR (Light Detection and Ranging). This involves a plane flying over the summit and measuring the time for a light-based signal to reflect off the summit back to the plane. The plane's position is known very accurately, and the time for the signal to return can be used to calculate the elevation. This measured a Columbia Crest summit elevation of 14,405.6 ft.

LiDAR has higher errors than using a ground-mounted GPS. Errors in flat terrain can be as low as +/-0.33 ft (USGS 2024), but signals are taken roughly every 3-6 ft (horizontal spacing) and can miss some features in terrain that's not flat, so errors on mountains can be higher.

Until now, the 14,411.0 ft measurement from 1998 has been the accepted elevation of Mount Rainier by the Land Surveyors Association of Washington. (Though, Rainier National Park still uses the 14,410 ft number from the 1956 survey; NPS 2024).

Starting in 2023, however, guides working for RMI (Rainier Mountaineering Institute) started reporting that it appeared that Columbia Crest had melted considerably over the previous few years, and that it now didn't look like it was the highest point in the summit area (as stated to EWG). A rocky point on the southwest edge of the crater rim appeared higher. Mount Rainier climbing guides go to the summit many times over the summer and return every summer, so they are uniquely qualified to make these observations.

In September, 2023, an attempt was made by the author (EWG) to climb and survey Mount Rainier, but the route had melted too much to proceed safely to the summit. Another surveying trip was planned for 2024.

# Methods

## Equipment

A Promark 220 differential GPS (dGPS) unit with an Ashtech antenna was used to take absolute elevation measurements. A mini prism tripod with a 1.0 ft antenna rod was used for measurements in the summit area (Columbia Crest, Summit 2 USGS monument location, and highest rock on the Southwest Rim). A 2.0 m antenna rod with an AdirPro prism tripod was used for measurements lower on the mountain (Muir, McClure, and Paradise USGS monuments).

A 10 arcminute 5x Sokkia Abney level and a 10 arcminute 1x Sokkia Abney level were used to take angular declination/inclination measurements between the Southwest Rim and Columbia Crest. These were used to corroborate the dGPS measurements.

## Measurements

Prior to the ground surveys, an initial estimate of the height difference between Columbia Crest and the Southwest Rim was made using photographic analysis. The Geopix surveying analysis software (Earl 2017) was used based on multiple pictures taken in July, 2023 by RMI guides from the top of Columbia Crest looking at the Southwest Rim. This software relies on identifying known peaks in the background of a picture, then entering coordinates, elevations, and pixel locations of background peaks and of the peak in question to calculate elevation. It outputs a relative height between the camera and the measured peak. The software has been validated on other peaks using theodolite, LiDAR, and dGPS ground surveys.

The initial ground measurements of Columbia Crest in 2024 were planned to be as close as possible to August 27th. The last ground measurement of Columbia Crest was taken on August 27, 1998, and thus a measurement close to August 27, 2024 would allow for fair comparison. Due to high wind, the 2024 measurement was delayed until August 28th.

At 8:30am on August 28th the Promark 220 dGPS was mounted on the highest point of glacial ice on Columbia Crest. The location was verified as the highest point on Columbia Crest using the Abney Levels. The dGPS unit was mounted on a tripod with a 1.0ft antenna rod, and data was logged for one hour (Fig. 3).

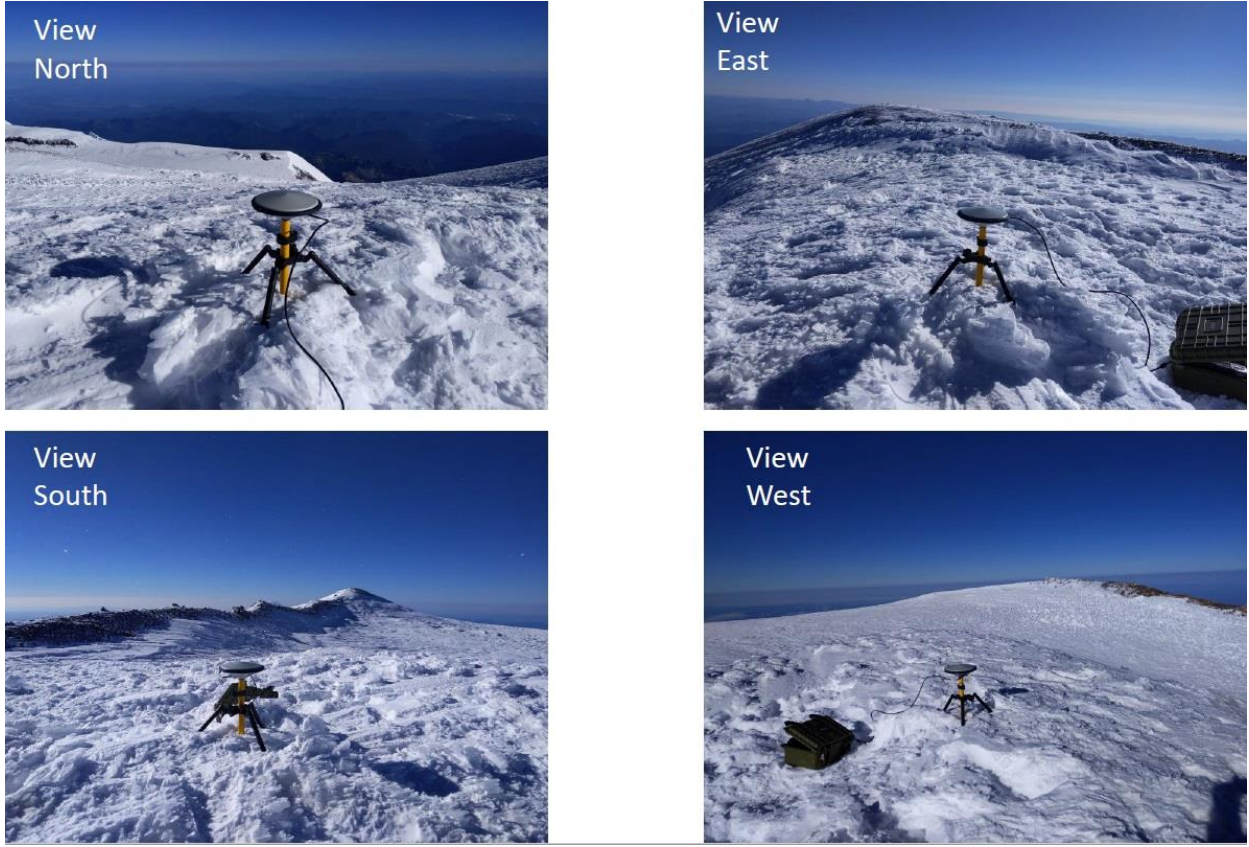

**Figure 3.** The dGPS unit mounted on Columbia Crest. Views look North, East, South, and West.

Figure 3. Views of the dGPS unit on Columbia Crest looking North, East, South, and West.

At 9:30am the unit was moved to the Southwest Rim and mounted on the top of the highest rock, a roughly 1ft x 2 ft x 3 ft boulder (Fig. 4). Abney level measurements indicated that this was the highest rock on the Southwest Rim and the highest point on the mountain. The unit was again mounted on the tripod using the 1.0ft antenna rod, and data was logged for one hour. Small rocks were used to anchor the tripod legs due to high wind.

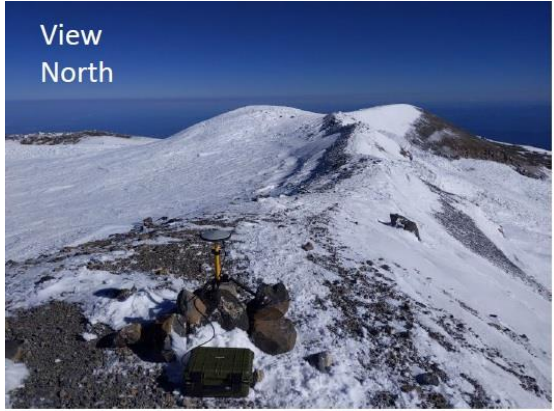 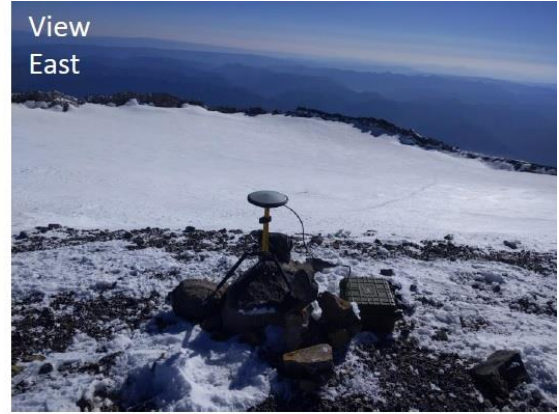
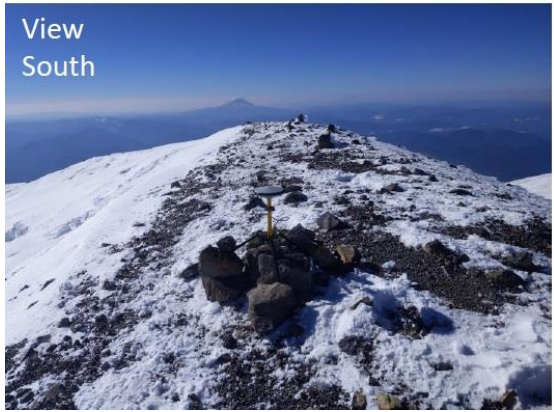 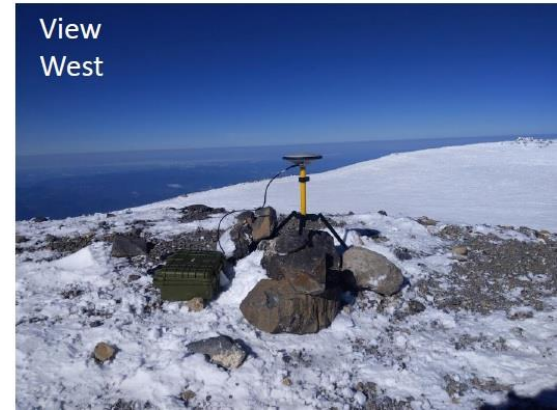

**Figure 4.** The dGPS mounted on the Southwest Rim highest boulder. Views look North, East, South, and West. Rocks were placed on the tripod legs to stabilize it against high winds.

The two Abney levels were used to measure angular declination from the highest point on the Southwest Rim down to the top of Columbia Crest, and angular inclination looking from the top of Columbia Crest up to the Southwest Rim highpoint.

Attempts were made to locate the Summit 2 USGS monument on the Southwest Rim and the Summit 1 USGS monument on the Northwest Rim, but they were unsuccessful. Anecdotal reports from guides indicate these USGS monuments have been missing for many years and were likely stolen. Measurements were planned for these monuments, but since the monuments were missing no measurements were taken of them.

To corroborate results, additional measurements were taken on September 21 and September 22, 2024.

On September 21, 2024, an attempt was made to locate the Summit 2 USGS monument using a metal detector, but was unsuccessful. This indicated that the underground metal pole below the monument may have been stolen. At 1pm, the Promark 220 was mounted at the approximate monument location on a tripod with the 1.0ft antenna rod, and data was logged for one hour. Fig. 5 shows pictures from the 2010 Summit 2 survey (left) and 2024 Summit 2 survey (right). Both views look North. The same summit boulder can be seen in both photos, shown with red arrow.

This surveyed location was next to the summit boulder that had previously been measured on August 28. A tape measure was used to measure the height of the top of the boulder above the surveyed location of the dGPS.

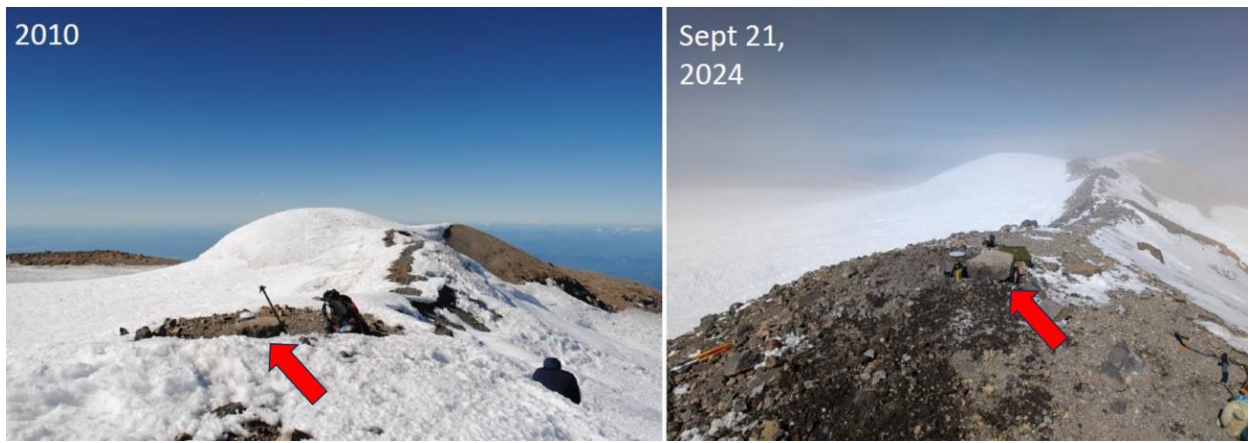

**Figure 5.** Summit 2 monument location measurements in 2010 (left) and 2024 (right). View looks North. Red arrow points to summit boulder in each photo.

Figure 5. View of the Summit 2 monument location measurements in 2010 and 2024 looking North.

On the descent, additional measurements were taken at the Muir, McClure, and Paradise USGS monument locations. At 2:30am Sept 22 the Promark 220 was mounted on the Muir monument location using the AdirPro tripod with a 2.0 m antenna rod. Data was logged for one hour. According to documentation from the 2010 LSAW survey, the monument was buried 0.9ft underground on the flat helipad at Camp Muir. The physical monument was not recovered but the antenna rod was mounted above the approximate monument location (Fig. 6).

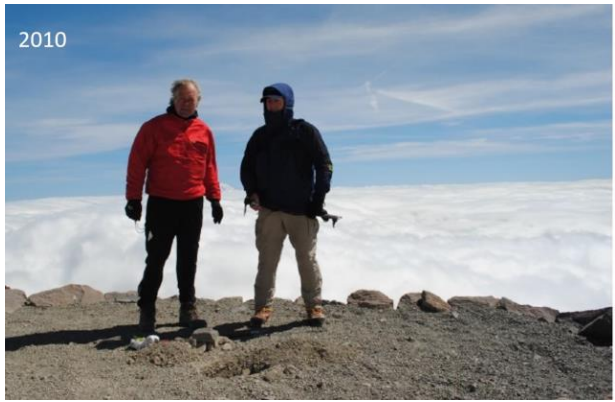 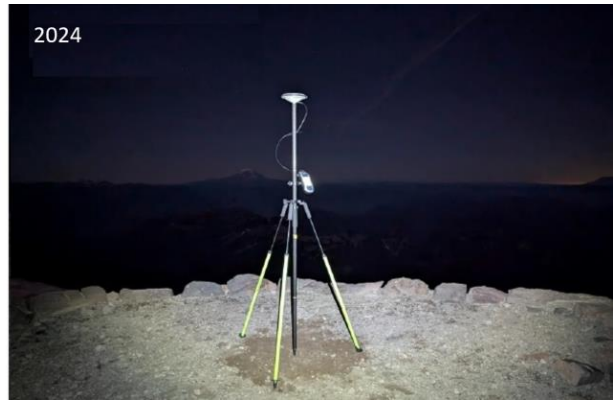

**Figure 6.** Muir monument measurements in 2010 (left) and 2024 (right). View looks South.

At 5:00am September 22 the Promark 220 was mounted on the McClure Rock USGS monument using the 1.0 ft antenna rod and the AdirPro tripod (Fig. 7). Data was logged for one hour.

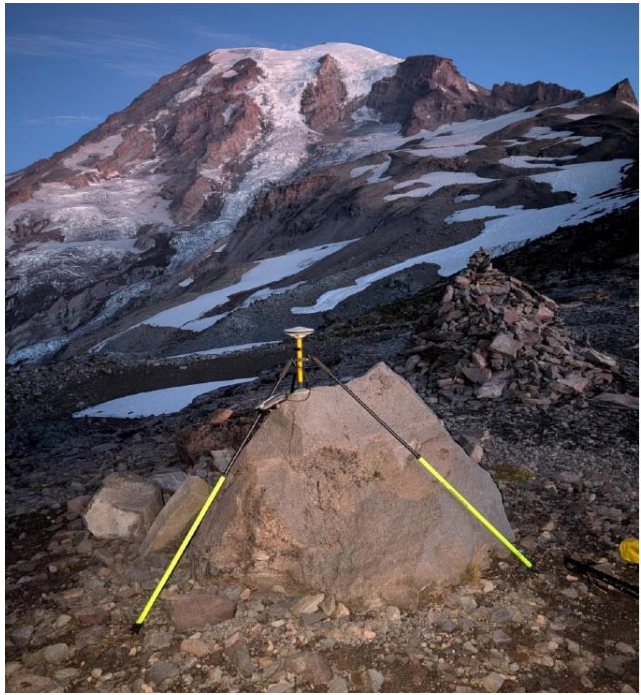 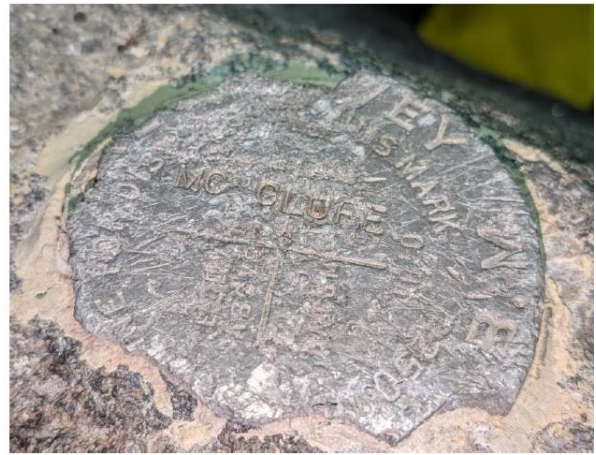

**Figure 7.** McClure monument measurements in 2024. View looks North with Mt Rainier in the background (left), and closeup view of monument (right)

At 9:00am September 22 the Promark 220 was mounted on the Paradise USGS monument location using the 2.0 m antenna rod. Documentation from previous surveys stated the monument was buried 0.4ft. The monument was not recovered, but the antenna rod was mounted above the approximate location of the monument (Fig. 8).

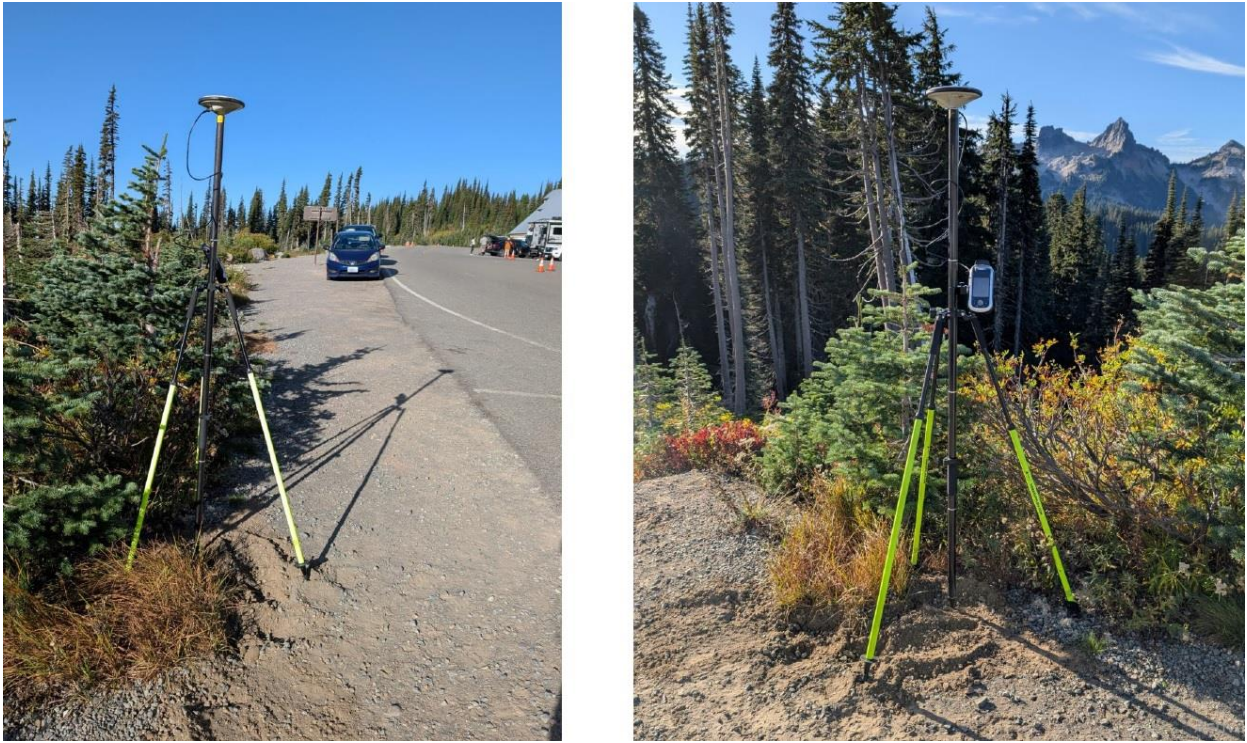

**Figure 8.** Paradise monument measurements 2024. Views look West (left) and South (right).

Figure 8. Views of the dGPS unit on the Paradise monument location looking West and South.

**Data Processing**

Data were processed with the Washington State Reference Network (WSRN 2024), Online Positioning User Service (OPUS 2024), and Canadian Spatial Reference System Precise Point Positioning processing (CSRS-PPP 2024). Final results will be reported as the WSRN results, though OPUS and CSRS-PPP gave consistent values. WSRN results were calculated using stations CPXF (Pack Forest), ENUM (Enumclaw), MRSD (Paradise), PKWD (Packwood), and ROKY (Graham).

Results were converted to NGVD29 datum using the National Geodetic Survey Coordinate Conversion and Transformation Tool (NCAT 2024), and all numbers are reported in NGVD29 to be consistent with the datum used by Rainier National Park and with those reported by previous surveys. Errors are reported as +/- two sigma (95% confidence interval).

LiDAR data from 2007 (LiDAR 2007) and 2022 (LiDAR 2022) were analyzed for the elevations of Columbia Crest and the Southwest Rim. Point cloud data was retrieved from the Washington Lidar Portal (WA Lidar Portal 2024) and the USGS National Map Downloader (National Map Downloader 2024) and analyzed using QGIS surveying software (QGIS 2024).

**Results**

A summary of all results can be found in Table 1. On August 28, Columbia Crest was measured with the Promark 220 at 14,389.2 ft +/- 0.1 ft (lat/lon 46.852950, -121.760572). The Southwest Rim highpoint was measured at 14,399.6 ft +/- 0.1 ft (lat/lon 46.851731, -121.760396). This means Columbia Crest was measured 21.8 ft shorter than the 1998 ground measurement. The Southwest Rim highpoint was measured 10.6 ft +/-0.2 ft taller than Columbia Crest, and is now the highest point on the mountain.

The Abney level measurements were averaged to find the angular declination from the Southwest Rim down to Columbia Crest for each device. The 5x Abney level measured a declination of 1.35 deg +/- 0.17 deg from the Southwest Rim down to Columbia Crest and the 1x Abney level measured a declination 1.25 deg +/- 0.17 deg from the Southwest Rim down to Columbia Crest.

Using the measured coordinates of the two locations from the GPS measurements, the locations were calculated to be 466 ft apart horizontal. Using trigonometry, this means the 5x Abney level measured the Southwest Rim 11.0 ft +/- 1.3 ft taller than Columbia Crest, and the 1x Abney level measured the Southwest Rim 10.2 ft +/- 1.3 ft taller than Columbia Crest.

These height differences are consistent with the dGPS measurements that the Southwest Rim is 10.6 ft +/- 0.2 ft taller than Columbia Crest.

The September 21 measurement of the Southwest Rim highpoint, taken by adding the boulder height to the Summit 2 measurement, was 14,399.6 ft +/- 0.1ft. This is the same result as measured on August 28.

Table 2 shows a comparison of the monument measurements from 2010 and 2024 for Summit 2, Muir, McClure, and Paradise USGS monument locations. The 2024 monument measurements were consistent with the 2010 measurements. (Note: these measurements are given in NAVD88 to be consistent with the originally-reported 2010 measurements).

**Table 2.** Comparison of USGS monument measurements from 2010 and 2024. These results are presented in NAVD88 to be consistent with how the 2010 results were reported.

| Monuments | 2010 (NAVD88) | 2024 (NAVD88) |
|---|---|---|
| McClure | 7389.78 ft | 7389.48 ft |
| Muir | 10086.90 ft | 10090.49 ft |
| Paradise | 5394.39 ft | 5393.78 ft |
| Summit 2 | 14407.24 ft | 14405.03 ft |

Geopix photographic analysis of the 2023 picture showed the Southwest Rim 8.0 ft +/- 0.5 ft taller than Columbia Crest on July 23, 2023.

The 2007 LiDAR data measured the Southwest Rim at 14,399.5 ft +/- 0.3 ft, which is within 0.1ft of the dGPS measurements. Because the dGPS measurement had errors of +/-0.1 ft and the LiDAR measurement has a nominal error of +/-0.3 ft (in flat terrain), these two measurements are consistent and within the error bounds of each other. Because LiDAR measurements are only taken roughly every 3-6 ft horizontal spacing, it is possible the LiDAR measurement did not hit the exact highest point of the summit boulder, and instead hit a slightly lower point on the boulder.

The 2007 LiDAR data measured Columbia Crest at 14,405.6 ft.

The 2022 LiDAR data measured Columbia Crest at 14,392.3 ft and the Southwest Rim highpoint at 14,398.7 ft. Interestingly, the height for the Southwest Rim from the 2022 Lidar measurement is 0.9 ft lower than the dGPS measurement. This difference is exactly the height measured with a tape measure of the summit rock above the surrounding dirt (0.9 ft). This could indicate that the 2022 LiDAR pass measured the dirt but missed the rock.

The Columbia Crest measurement was shorter in 2024 than in 2022 and 2007, which is consistent with the climbing guide-reported observations that Columbia Crest is shrinking.

To understand how the elevation of Columbia Crest has changed over time, Table 1 shows all elevation measurements of Columbia Crest and the Southwest Rim from 1841 to present, including the exact date and measurement error (+/- two sigma, 95% confidence interval) when known. Table 1 shows that the early measurements in the 1800s were prone to high errors, but measurements were more accurate by 1914 and 1956.

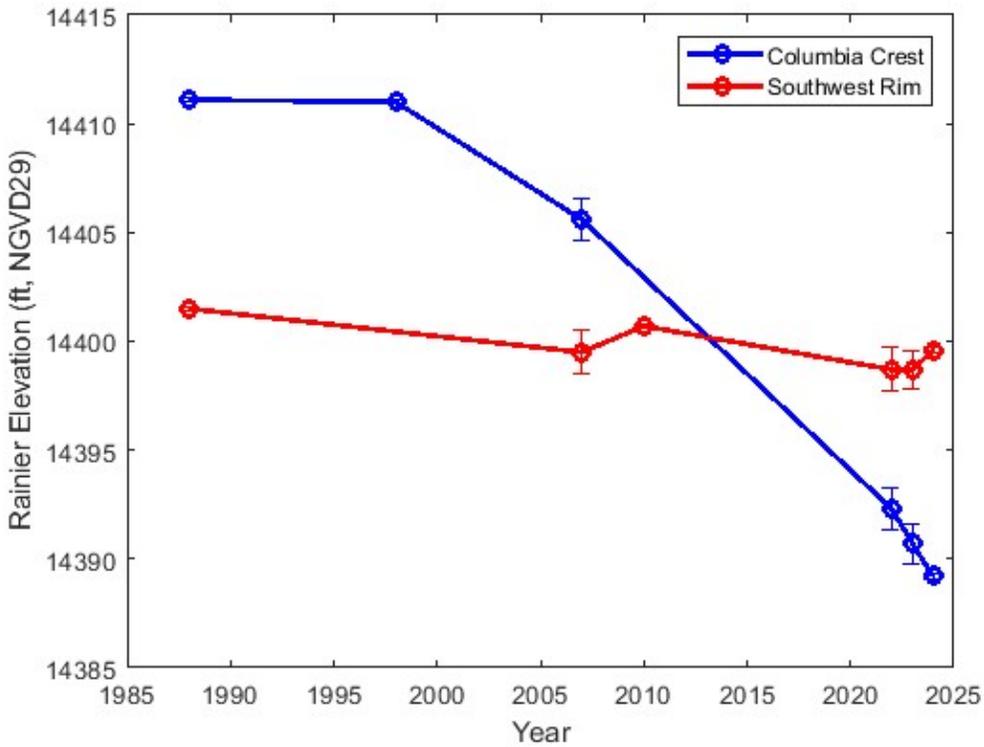

**Figure 9.** Columbia Crest and Southwest Rim elevations over time from 1988 to 2024.

Figure 9. Graph showing the elevations of Columbia Crest and the Southwest rim over time between 1988 and 2024.

Columbia Crest has been losing elevation since 1998. In approximately 2014, Columbia Crest became shorter than the Southwest Rim, and the Southwest Rim thus became the location of the summit of Mt Rainier (Fig. 9).

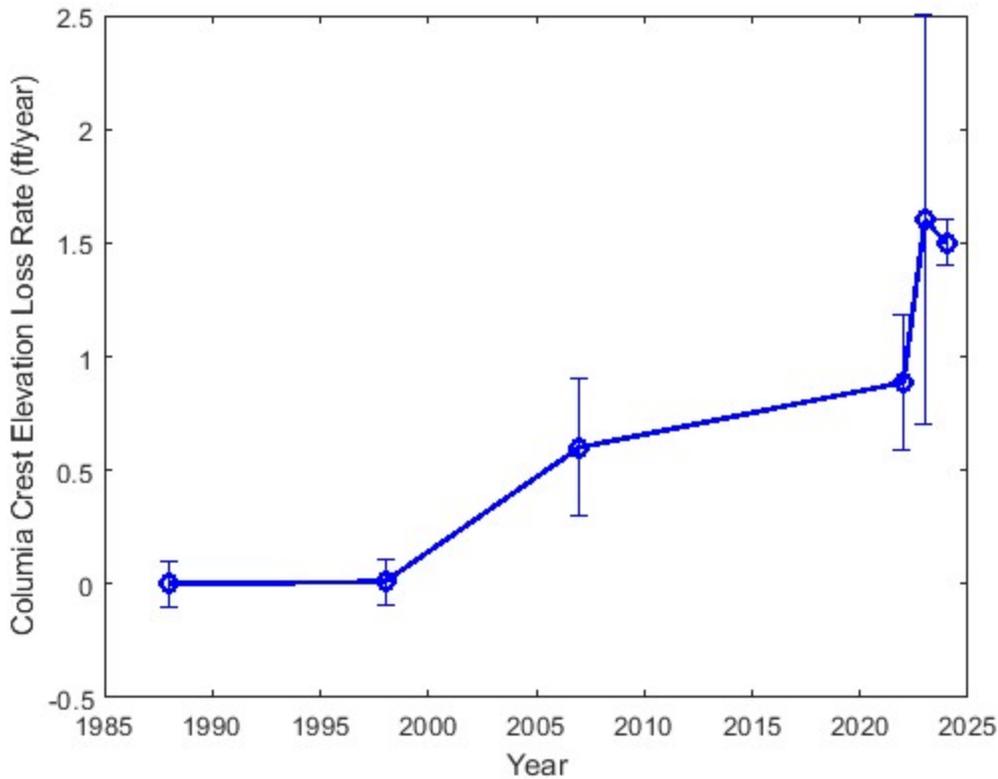

**Figure 10.** Plot of elevation loss rate (ft/year) of Columbia Crest between 1988 and 2024.

Figure 10. Graph showing the elevations loss rate per year of Columbia Crest between 1988 and 2024.

Before 1998, the elevation of Columbia Crest was relatively constant. Between 1998 and 2007, Columbia Crest lost on average about 0.7 ft per year. This rate increased to 0.9 ft/year between 2007-2022, then 1.6 ft/year between 2022-2023, then 1.5 ft/year between 2023-2024. The rate of elevation loss during the past two years is the same within the error bounds of the measurements, so it is unclear if the rate is increasing, decreasing, or staying constant (Fig. 10).

To visualize the extent of melting of Columbia Crest, Figure 11 shows the LiDAR point cloud data for the western crater rim from 2007 and 2022. The upper region is Columbia Crest and the lower region is the Southwest Rim. Pixel colors represent elevation, ranging from 14,390 ft (light blue) to over 14,400 ft (dark red). In the 2007 image Columbia Crest is clearly larger and taller than in the 2022 image. It is taller than the Southwest Rim in the 2007 image and shorter in the 2022 image.

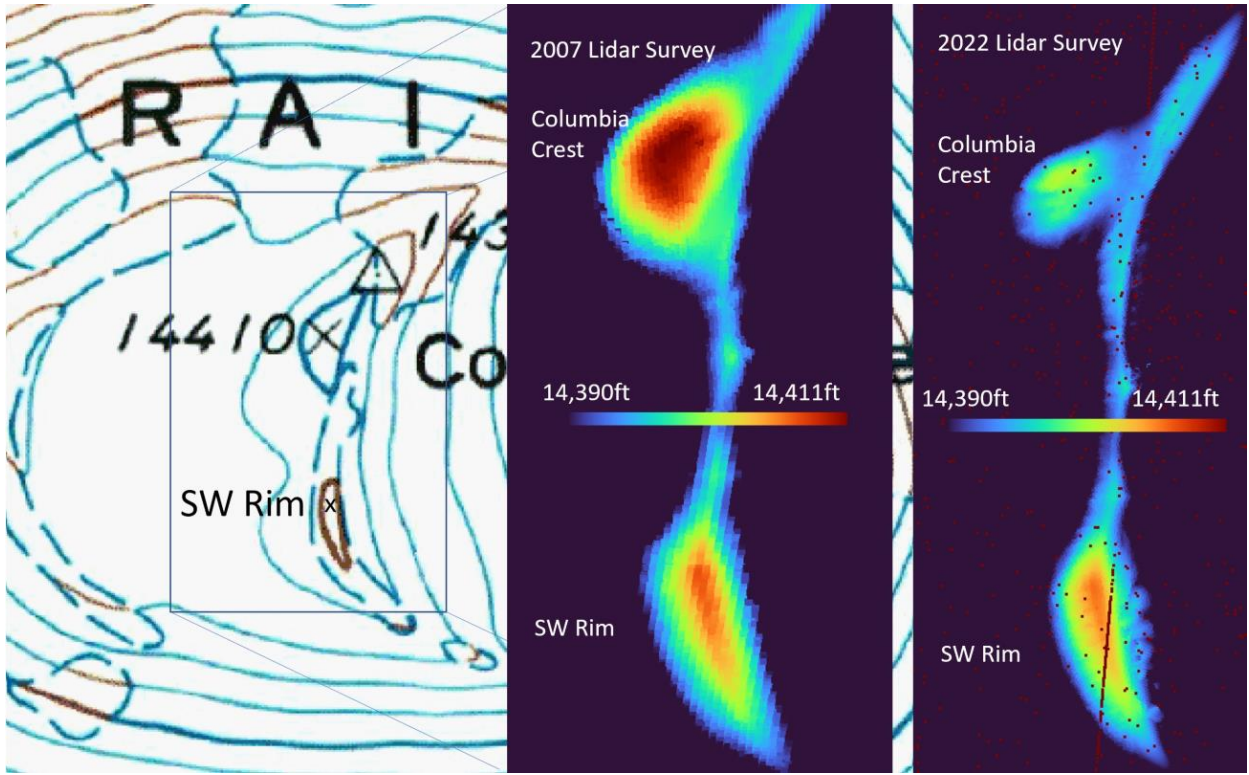

**Figure 11**. 2007 and 2022 Lidar point cloud data showing Columbia Crest taller than the Southwest Rim in 2007 and shorter in 2022.

Figure 11. Picture showing the Lidar point cloud data from 2007 showing Columbia Crest taller than the Southwest Rim and from 2022 showing Columbia Crest shorter.

Figure 12 shows the view from June 27, 2009 looking from the SW Rim towards Columbia Crest (Arhuber 2009). Figure 13 shows the same view from 2024. Columbia Crest is clearly much larger and taller in 2009 than in 2024.

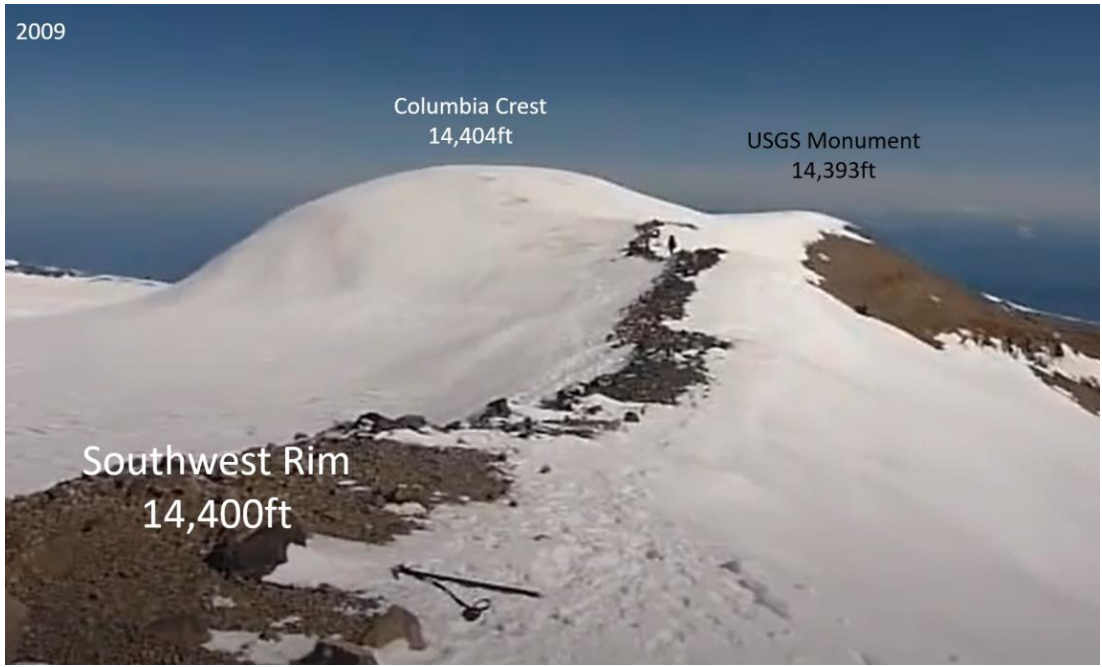

**Figure 12**. The view from the Southwest Rim looking towards Columbia Crest in 2009 (Arhuber 2009; elevation interpolated from Fig. 3).

Figure 12. Photograph taken in 2009 from the Southwest Rim looking towards Columbia Crest

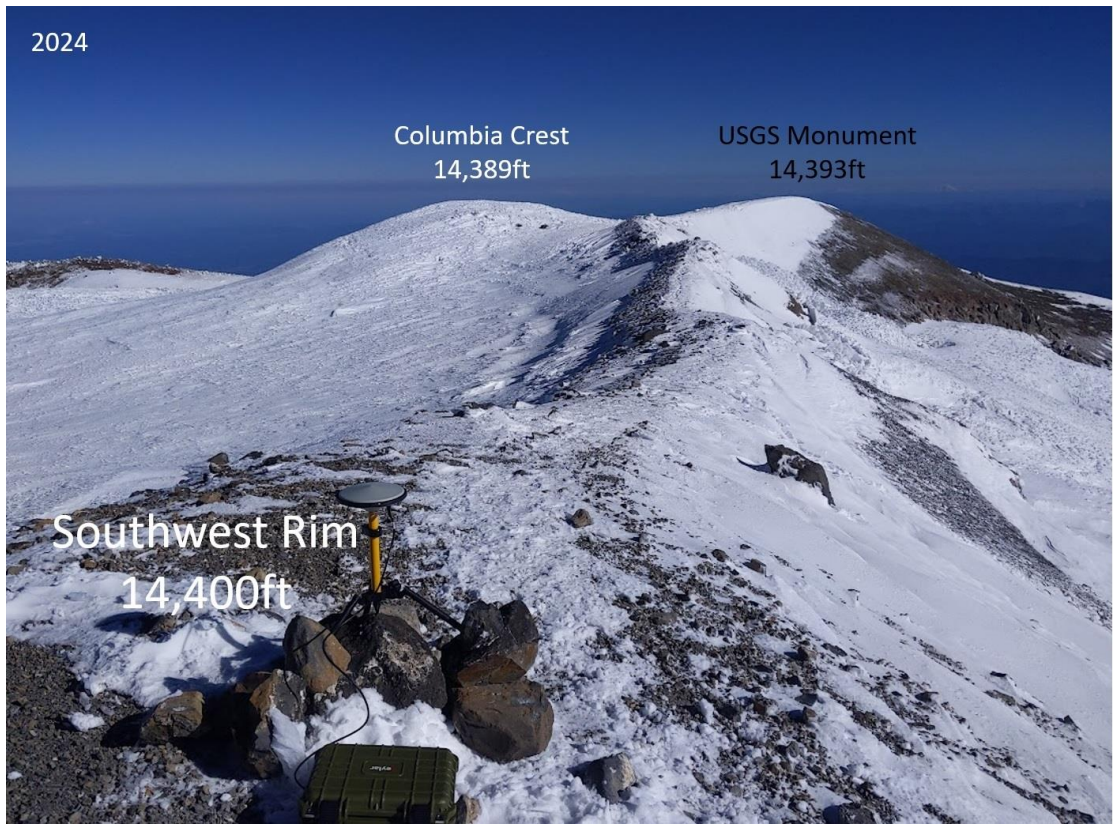

**Figure 13**. The view from the SW Rim looking towards Columbia Crest in 2024.

Figure 13. Photograph taken in 2024 from the Southwest Rim looking towards Columbia Crest

## Discussion

Four independent measurement methods (dGPS, Abney level, Lidar, and photographic) are all in agreement that Columbia Crest is no longer the highest point on Mount Rainier. The current location of the summit of Mount Rainier is the highest rock on the Southwest Rim.

The elevation of the highest point on Mount Rainier on the Southwest Rim was measured at 14,399.6 ft NGVD29 by dGPS on both August 28 and September 21. This value is consistent with the LiDAR measurements from both 2007 and 2022.

The Southwest Rim has been measured between 14,401.5 ft (1988) and 14,399.6 ft (2024). It's unclear if the height reduction was from snow melt, erosion, or some other reason. The 2024 measurement was on rock. The Summit 2 USGS monument has been stolen multiple times, so it's possible the different heights are the results of measurements taken at slightly different locations.

dGPS measurements were corroborated by USGS monument measurements on the mountain, further increasing confidence in the summit elevation results. The McClure and Paradise monument measurements matched very closely. The Summit 2 measurements differed by slightly more, and this is likely because the monument was stolen between measurements so slightly different locations were measured. The Muir monument measurements differed slightly more, and this is possibly due to the Muir helipad being modified since 2010 (as is evident with the rocks outlining the helipad being different between 2010 and 2024).

Based on data points from 1988, 1998, 2007, 2022, 2023, and 2024, Columbia Crest is losing elevation. Elevation loss has occurred between every pair of successive measurements since 1998, and the rate of loss is accelerating. The elevation loss began in approximately the early 2000s and has continued in each subsequent measurement. The current rate of loss is approximately 1.5ft per year.

In approximately 2014, Columbia Crest lost enough elevation that it was no longer the highest point on Mount Rainier, and instead the Southwest Rim became the highest point on the mountain.

## Conclusion

The Columbia Crest icecap that used to be the highpoint of Mount Rainier has lost 21.8 ft of elevation since 1998. Columbia Crest is no longer highest point on Mount Rainier, and instead the Southwest Rim is the highest point at 14,399.6ft NGVD29 (14,406.2ft NAVD88).

Washington Lidar Portal, 2024. https://lidarportal.dnr.wa.gov/

USGS National Map Downloader, 2024. https://apps.nationalmap.gov/downloader/